\documentclass[fleqn,usenatbib]{mnras}

\usepackage{newtxtext,newtxmath}

\usepackage[T1]{fontenc}

\DeclareRobustCommand{\VAN}[3]{#2}
\let\VANthebibliography\thebibliography
\def\thebibliography{\DeclareRobustCommand{\VAN}[3]{##3}\VANthebibliography}

\usepackage{subfig}

\usepackage{graphicx}	
\usepackage{amsmath}	
\usepackage{newtxtext,newtxmath}	
\usepackage{multirow}
\usepackage{tabularx}
\usepackage{arydshln}

\title[Towards a coupled activity-fallback model for comets I]{Towards a coupled activity-fallback model for comets I) 67P/Churyumov-Gerasimenko's water activity from a microphysics model with dynamic dust-depth}

\author[N. Attree et al.]{
N.~Attree$^{1}$\thanks{E-mail: attree@iaa.csic.es},
Y.~Skorov$^{2,3}$,
P.~Gutiérrez$^{1}$,
O.~Groussin$^{4}$,
A.~Jindal$^{5}$,
R.~Marschall$^{6}$,
S.P.D.~Birch$^{5}$
\\
$^{1}$Instituto de Astrofísica de Andalucía (CSIC), Glorieta de la
Astronomía s/n. 18008 Granada, Spain\\
$^{2}$Institut f\"ur Geophysik und Extraterrestrische Physik, Technische Universit\"at Braunschweig, Mendelssohnstr. 3, 38106 Braunschweig, Germany\\
$^{3}$Max-Planck-Institut f\"ur Sonnensystemforschung, Justus-von-Liebig-Weg 3, 37077, G\"ottingen, Germany\\
$^{4}$Aix Marseille Univ, CNRS, CNES, Laboratoire d’Astrophysique de Marseille, Marseille, France.\\
$^{5}$Department of Earth, Environmental, and Planetary Sciences, Brown University, Providence, RI, USA.\\
$^{6}$Physikalisches Institut, Sidlerstr. 5, University of Bern, CH-3012, Bern, Switzerland\\
}

\date{Accepted XXX. Received YYY; in original form ZZZ}

\pubyear{2026}

\begin{document}
\label{firstpage}
\pagerange{\pageref{firstpage}--\pageref{lastpage}}
\maketitle

\begin{abstract}

We apply a two-layer thermophysical model, with water sublimating from below a dust-crust with dynamically varying depth, to simulate the water outgassing rate, dust-ejections, and non-gravitational forces and torques of comet 67P/Churyumov-Gerasimenko. The dust-depth is determined at each time-step by the balance of dehydration and subsurface pressure, leading to alternating ejections and crust build-up. This model is the first stage of implementing a fully coupled activity-fallback model that self-consistently incorporates dust-fallback into the ejection depth calculation, but already a number of conclusions can be drawn from these results. The two-layer model with only dehydration and ejections is able to reproduce much of the Rosetta outgassing, dust-ejection data and non-gravitational dynamics, particularly in the comet's more primitive southern terrain. Best agreement is found for nucleus material made up of small aggregates ($d_{agg}\sim1$ mm) with not-too-high dust-to-ice ratios ($\delta\sim2-8$) and low average strengths ($\lesssim1$ Pa) that decreases with depth. Disagreements between the observations and model are greatest for the northern-facing and dust-covered terrains, suggesting the importance of surface modification by fallback and material processing here, which will be the subject of investigation in the next paper. 
\end{abstract}



\section{Introduction}
\label{sec:introduction}

Explaining the activity of comets, i.e.~the ejection of gas and dust-particles from their surfaces by the action of sublimating ices, remains a challenge for numerical and conceptual models. The scarcity of exposed ice on cometary surfaces (e.g.~see \citealp{Keller1987, Fornasier2023}, etc.~) suggests that ice sublimates from below a dust-crust \citep{Kuehrt1994} or from within aggregates composed of a mixture of icy and refractory grains \citep{Fulle2019}. Overcoming the tensile strength that holds these grains together within aggregates, or that binds aggregates together into the crust, has previously been considered extremely difficult \citep{Kuehrt1994, Jewitt2019}, but recent work suggests it is possible by water-ice sublimation \citep{Gundlach.2020, Skorov2024b, Attree2025}. The aggregate structure plays a key role here in reducing the effective tensile strength from the $\sim$kilo-pascals of mono-disperse micron-sized grains to the $\sim$pascal level \citep{Blum2017}.

When simulating the transport of heat into a cometary surface, and the resulting sublimation and pressure build-up, various approaches have been taken. The simplest approach is to solve for the surface-energy-balance, assuming an icy surface, and merely scale the outgassing fluxes down by some `effective active fraction', or EAF (referred to as Model A in \citealp{Keller2015}). This approach has proved useful for deriving empirical activity curves for comparison with measured cometary outgassing rates, non-gravitational accelerations (NGAs) and torques (NGTs) \citep{Attree2018, Attree2023, Attree2024}, but does not explain the physical processes behind the EAF, nor simulate the dust ejections.

At the opposite end of the complexity scale are discretised numerical solutions that solve the coupled heat- and gas-flows (e.g.~\citealp{Davidsson2021MNRAS, Davidsson2022, Zivithal2025}), as well as those that solve the heat-flow but approximate the gas-flow (e.g.~\citealp{Gundlach.2020, bischoff2023, Attree2024b, Attree2025}). These provide much more physical insight, at the cost of increased computational demand. Intermediate between these approaches are methods that approximate the cometary interior by two layers (referred to as Model B in \citealp{Keller2015, Skorov2024b}, and also adopted in \citealp{Jindal2022}). These layers are typically taken as a dry dust-crust overlying a dust-ice mixture below, where the sublimation is assumed to occur at the boundary between the layers, and the gas flows through the crust according to its known permeability \citep{Guettler23, Macher2024}. A later enhancement to this model is to then compare the sublimation pressure at the layer boundary to some material strength in order to calculate dust ejections. This kind of model was applied to comets in general by \citet{Xin2025a} and then to comet 67P/Churyumov-Gerasimenko (hereafter, 67P) in particular \citep{Xin2025b}, although a full comparison with 67P's outgassing rates, NGA and NGT was not carried out. 

67P represents a particularly relevant target for activity modelling due to the wealth of data returned on it by ESA's Rosetta mission, while its complex shape and seasons present an interesting challenge. \citet{Keller2015}, and more recently \citet{Groussin2025}, showed the importance of these seasonal insolation variations in setting the contrasting thermal environments of its northern and southern hemispheres; the former experiencing a long, cold summer at aphelion, and the latter, a short but intense heating around perihelion. These seasonal trends likely shape the morphological appearance of 67P's nucleus, with the north dominated by smooth, dusty fallback deposits, and the south by exposed consolidated material \citep{ElMaarry15, Birch17, Thomas2018}. A global transport of dust, ejected from the active south and falling back on the less-active north, is implied \citep{Keller2017}, and several attempts have been made to model this \citep{Lai2017, Kloos2025, Jindal2026}. Returning to the modelling of 67P's outgassing rates, meanwhile, several authors have shown that varying material properties are needed to fit the steeply rising water-production around perihelion. This can take the form of a changing EAF \citep{Attree2024} in surface-energy-balance models, or a changing depth of dust-crust in more complex models (\citealp{Skorov2020, Davidsson2022}, with the latter also varying other physical parameters). 

Taken together with the above observations on 67P's geomorphology and dust-transport, there is an implied connection between surface dust-cover and activity-level that can only be fully understood by taking both into account in a coupled model. The work of \citet{Kloos2025} went some way towards achieving this by scaling a modelled dust-ejection flux by the water outgassing from a two-layer thermal model, and tracking the resulting fallback distribution. However, the subsequent outgassing activity was not then modified by the changing dust depth due to the dehydration and fallback, i.e.~the coupling was only one-way, forwards from the thermal model to the ejection/fallback calculation. The relatively simple and fast two-layer model with ejections offers the opportunity to include this coupling in the other direction, back from a dynamically adjusting dust depth to the water outgassing and ejection rate, in order to simulate the coupled system more completely. This is the goal of our present work: a fully-coupled model linking a thermophysical description of outgassing and ejections from below a dust crust \citep{Xin2025a, Xin2025b} with a detailed mapping of the fallback particle trajectories \citep{Jindal2026} and their effects on the depth of this crust. Transport of volatile ices within the dust will also be considered in the future. In this way, we hope to eventually simultaneously constrain cometary outgassing, non-gravitational dynamics, dust ejections, and surface geomorphology.

In the current paper we present the first stage of this model: testing the two-layer model with ejections on 67P, and comparing the results to all the related Rosetta data, as well as those of a more complex thermophysical model \citep{Attree2025}. In Section \ref{sec:method}, we give the details of the model and describe how it is implemented here, before describing the results for various input parameters in Section \ref{sec:results}. In Section \ref{sec:discussion}, we discuss these results in the context of Rosetta's measurements of 67P's water-outgassing, dust-fluxes, NGAs, and NGTs. Conclusions and perspectives for the inclusion of the fallback mapping are drawn in Section \ref{sec:conclusions}.

\section{Method}
\label{sec:method}

We re-introduce the two-layer model of \citet{Xin2025a} and \citet{Xin2025b}, wherein water-ice is assumed to sublimate either directly from the surface, or from below a dry dust-crust composed of millimetre to centimetre-sized aggregates of micron-sized grains (i.e.~a hierarchical material with two levels of porosity: \citealp{Skorov.2012}). We apply the model to every facet of the 67P shape model ({\it SHAP7} with 125k facets; \citealp{Preusker2017}) using the same procedure as \citet{Attree2024}. The thickness of the dust-crust, $d$, is dynamically adjusted at each time-step according to

\begin{equation}
    d = \rm{Dehydration} - \rm{Erosion},
    \label{eq:d}
\end{equation}
where the first term on the right-hand-side is given by
\begin{equation}
    \rm{Dehydration} = \frac{Z(1+\delta)}{\rho} dt.
    \label{eq:dehydration}
\end{equation}
Here $Z$ is the water sublimation-rate, $\delta$ the bulk dust-to-ice mass ratio and $\rho=537.8$ kg/m$^{3}$ the bulk density \citep{Preusker2017}, and $dt$ is the time-step. The water sublimation-rate depends on the ice temperature, $T_{ice}$, and location, but its general form is
\begin{equation}
Z(T_{ice}) = P_{sat} \sqrt{\frac{M}{2\pi R T_{ice}}},
\label{eq:z}
\end{equation}
where the saturation vapour pressure is $P_{sat} = A e^{-B/T_{ice}}$, with molar mass $M = 0.018$ kg, gas constant $R = 8.3144598$ J K$^{-1}$ mol$^{-1}$, and sublimation constants $A=3.56\times10^{12}$ Pa and $B = 6162$ K for water \citep{FanaleSalvail}. 

Whenever the dust-crust is less than one aggregate thick ($d<2r_{agg}$), sublimation is assumed to be happening at the surface, and a standard surface energy-balance equation is solved to obtain Z($T=T_{surf}=T_{ice})$:
\begin{equation}
\frac{(1-\alpha)F_{\rm sun}\cos z}{r_{\rm h}^2} +SH= e \sigma T^4 + LZ(T),
\label{eq:SEB}
\end{equation} with albedo and emissivity $\alpha=0.04$ and $e=0.968$, solar constant $F_{\rm sun}=1370$ W/m$^2$, latent heat of sublimation of water ice $L=2.75\times10^6$ J/Kg, time-dependent heliocentric distance $r_h$, and facet zenith angle $z$, and self-heating $SH$. The self-heating term is calculated as in \citet{Attree2024} and \citet{Groussin2025} using pre-calculated facet view factors and black-body surface temperatures. This means that it is a maximum estimate, particularly when ice is on the surface and would be significantly cooling it by sublimation.

When dehydration brings the water sublimation-front below the first aggregate ($d>2r_{agg}$), the two-layer model of \citet{Skorov2024b} is used (their Model B). Here, two coupled equations are solved for two separate temperatures, $T_{surf}$ and $T_{ice}$, at the top surface and sublimation-front at depth d, respectively:
\begin{equation}
\begin{aligned}
\frac{(1-\alpha)F_{\rm sun}\cos z}{r_{\rm h}^2} +SH=e \sigma T_{surf}^4 + \lambda(T)\frac{T_{surf} - T_{ice}}{d}, \\
\lambda(T)\frac{T_{surf} - T_{ice}}{d} - \lambda(T)\frac{T_{ice} - T_{in}}{d_{in}} = \Psi L Z(T_{ice}),
\label{eq:2-layer}
\end{aligned}
\end{equation}
where the second term in the second equation represents the temperature gradient into the nucleus interior at an assumed $T_{in}=30$ K below $d_{in}=0.1$ m. In these equations, thermal conductivity, $\lambda$, and gas permeability, $\Psi$, are determined by material and structural properties such as the mean free path for gas molecules and photons, given by \citep{Macher2024, Guettler23}
\begin{equation}
\rm{MFP} = \frac{4\epsilon}{3(1-\epsilon)}r_{agg},
\label{eq:MFP}
\end{equation}
for aggregates of radius $r_{agg}$ and an effective aggregate-packing porosity of $\epsilon$ (i.e.~the top level porosity of the hierarchically structured material). Gas-permeability is then
\begin{equation}
\Psi = \frac{1}{1+d/b}, \qquad b = \frac{3}{4}\rm{MFP},
\label{eq:eta}
\end{equation}
using the half-transmission thickness, or $b$ formalism (\citealp{Gundlach.2011b}, and see further discussion in \citealp{Attree2024b}).

Thermal conductivity, $\lambda=\lambda_{net}+\lambda_{rad}$, is the sum of the network conductivity between contacting pebbles, which is material dependent and calculated as in \citet{Gundlach.2020, bischoff2023} and \citet{Attree2025}, and a temperature dependent radiative component given by
\begin{equation}
\lambda_{rad} = 8\epsilon\sigma T^{3}  \rm{MFP}.
\label{eq:lambda}
\end{equation}
with Stefan–Boltzmann constant $\sigma$.

For a given dust-to-ice ratio, aggregate size, and packing porosity ($\delta$, $r_{agg}$, and $\epsilon$), equations \ref{eq:2-layer} are then solved for a range of total energy inputs (solar plus self-heating) and dust-depths, creating a fast lookup table which can be applied to every facet of the shape model at each time-step to calculate $T_{surf},T_{ice}$ and $Z$.

When water is sublimating from below a dust crust ($d>2r_{agg}$), the vapour and ice are assumed to be in thermodynamic equilibrium, so that the pressure equals the saturation pressure, $P_{sat}$, at the ice temperature, $T_{ice}$. This pressure is compared to the material tensile strength plus the pressure from its weight, equal to $\rho gd$ for 67P's average surface gravity of $g=2\times10^{-4}$ m s$^{-2}$. If this combined strength is overcome then the crust is considered ejected, the depth $d$ is reset to zero, and the ejected mass is added to the tracking of dust-flux, assuming it is dry and contains no water for simplicity,
\begin{equation}
     Z_{dust} = \frac{\rho\delta(1+\delta)\rm{Erosion}}{dt}, \qquad \rm{Erosion} = d.
    \label{eq:Erosion}
\end{equation}
While the assumption of dry dust will be revisited in the future, we should note here that this binary, on or off, ejection model directly sets the size range of ejected particles (equal to $d$) as well as the activity behaviour, i.e.~a rather sharp cut-off between active and inactive facets at a particular time.

At each time-step the outgassing force is calculated with 
\begin{equation}
F = \eta  Z(T) \sqrt{\frac{8RT_{surf}}{\pi M}},
\label{eq:force}
\end{equation}
where the term under the square root represents the gas thermal speed. Here it is assumed that gas molecules are heating to the surface temperature whilst passing through the dust-crust \citep{Skorov2024a}. The momentum transfer coefficient is set to $\eta=0.7$ as per \citet{Attree2018}. The direction of the force vector is assumed to be along the facet normal, and torque per facet is calculated by the cross product with facet location. Momentum transfer, gas-heating and collimation involve complex interactions between the gas-flow and the structure of the dust-crust and surface. These are beyond the scope of the current, global analysis and will be the subject of a separate work \citep{Skorov2026}.

The total output fluxes at each time-step are found by summing the relevant quantities over all $i$ facets, taking into account their surface areas $a_{i}$:
\begin{equation}
X=\sum_{i} a_{i}x_{i},
\label{eq:total}
\end{equation}
for, e.g.~$x_{i}=Z_{water, i}$ and $Z_{dust, i}$. Mean values of these totals over 67P's rotation period (12.4 hrs) are then typically displayed in the following plots. Total force is used to calculate acceleration curves with 67P's known mass, and total torque is integrated with the Euler equations to evaluate the evolution in spin-rate and pole-direction, following the method in \citet{Attree2024}. The model is started from an initial condition of $d=0$ m everywhere, and run over two repeated orbits, after-which repeating periodic behaviour is observed in the active facets (facets that never experience ejections undergo a secular increase in crust-depth).

\begin{figure}
\resizebox{\hsize}{!}{\includegraphics{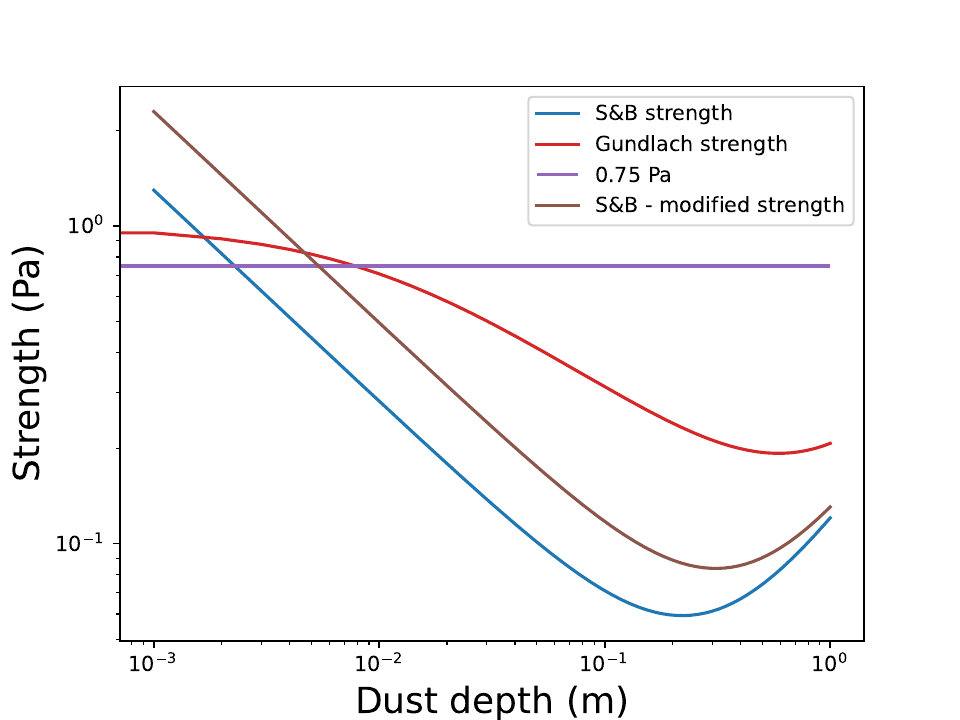}}
\caption{Material tensile strength plus weight (responsible for the up-turn after a few tens of centimetres)} as a function of depth for two different depth-dependent scaling laws from \citet{Skorov.2012} and \citet{Gundlach.2020}. Also shown is a fixed tensile strength of 0.75 Pa and a modified version of \citet{Skorov.2012} (see Sect.\ref{sec:discussion}).
\label{plot_strength}
\end{figure}

With the spatial and temporal resolutions and input conditions (insolation and self-heating and shadowing) set, the only free parameters in the model are:
\begin{enumerate}
    \item The aggregate-packing porosity, which should be around the packing porosity for loosely packed spheres, $\epsilon\approx0.3-0.4$, to ensure the total measured 67P porosity when combined with the internal aggregate porosity in the hierarchical structure (see discussion in \citealp{Attree2024b} Sect. 2.4).
    \item The dust-to-ice mass ratio, $\delta$, which controls the rate of dehydration only (Eqn.~\ref{eq:dehydration}).
    \item The size of the aggregates, $r_{agg}$, which determines the gas permeability (Eqn.~\ref{eq:eta}) and the radiative part of the thermal conductivity (Eqn.~\ref{eq:lambda}) through the mean free path.
    \item The material strength, which determines the erosion through comparison with the saturation pressure, $P_{sat}$.
\end{enumerate}
We constrain the packing porosity as above and, for the next two parameters,  test values corresponding to reasonable upper and lower limits for water-content ($\delta=2,8$) and aggregate size ($d_{agg}=1$ mm, 1 cm), based on 67P data \citep{Choukroun2020, Blum2017}. Material strength is still poorly constrained (see discussion in \citealp{Attree2024b}) so, rather than linking it directly to aggregate size by theory \citep{Skorov.2012}, we test a number of different strength laws. Figure \ref{plot_strength} shows the depth-dependencies used, corresponding to the size-defects law used in \citet{Gundlach.2020} and an equating of depth with increasing particle size, with a tensile strength given by \citet{Skorov.2012}. We also test various fixed tensile strengths within the same range. It can be seen from Fig.~\ref{plot_strength} that tensile strength dominates over weight for all ejected particles smaller than $\sim20$ cm.

\section{Results}
\label{sec:results}

The model is run for the range of input parameters shown in Table \ref{tab:results}. A similar pattern is seen for all parameter sets: gas emission from below a dust-crust, followed by increasing insolation triggering ejections and leading to intermittent crust build-up and surface emission, followed by a dying-out of activity with decreasing insolation, and a slowly increasing crust thickness. Average dust-depth is then correlated with heliocentric distance, while relative gas emission is strongly anti-correlated. This pattern repeats itself after the first perihelion, with dust depths at the following aphelions generally reaching centimetres to tens of centimetres for the millimetre- to centimetre-sized aggregates, respectively.

\begin{table}
\caption{Model input parameters (aggregate size, $d_{agg}$, dust-to-ice mass ratio, and the strength law used: see FIg.~\ref{plot_strength}), and the outputted reduced $\chi^{2}$ compared to Rosetta observations (see \citealp{Attree2024} for $\chi^{2}$ calculation and weighting). See text for details of the Notes.}
\label{tab:results}
\begin{tabular}{ccccl}
\hline
$d_{agg}$ & D2I & Strength & $\chi^{2}$ & Notes \\
\hline
1 cm & 2 & S\&B & 18.05 & - \\
1 mm & 2 & S\&B & 7.67 & - \\
1 mm & 8 & S\&B & 7.51 & - \\
1 mm & 2 & S\&B & 13.15 & Porosity 0.5 \\
1 mm & 2 & S\&B & 6.46 & Porosity 0.3 \\
1 mm & 2 & S\&B & 7.41 & Surface model transition \\
1 mm & 2 & Modified S\&B & 1.08 & Heterogeneous \\
1 mm & 2 & Gundlach & 2.54 & - \\
1 mm & 2 & 0.75 Pa & 2.25 & - \\

\hline
\end{tabular}
\end{table}

\begin{figure*}
\subfloat{\includegraphics[scale=0.5]{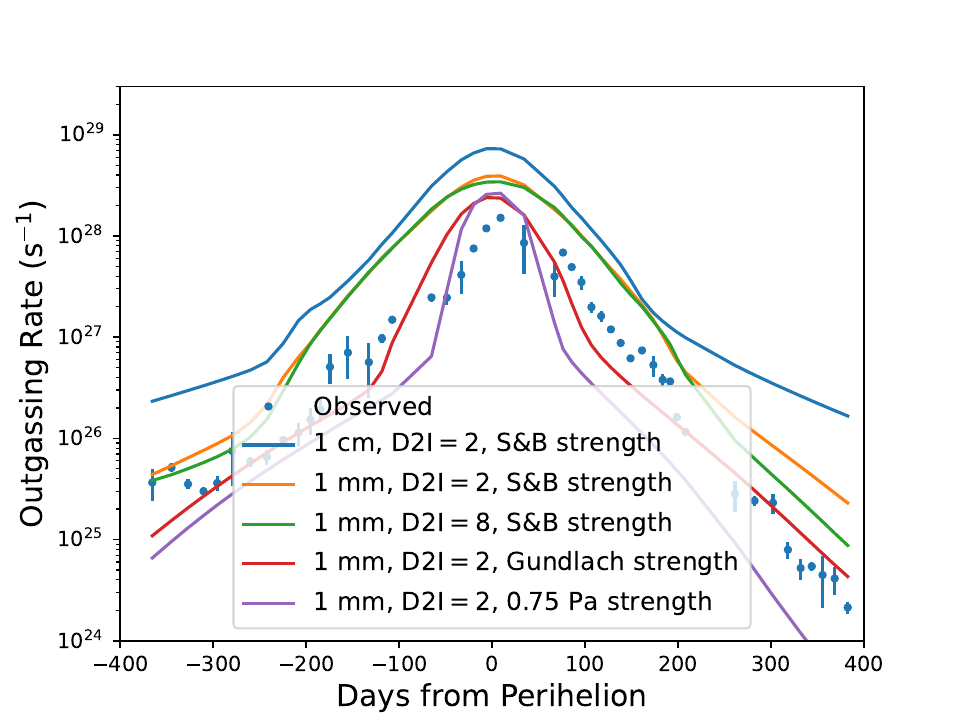}}%
\qquad
\subfloat{\includegraphics[scale=0.5]{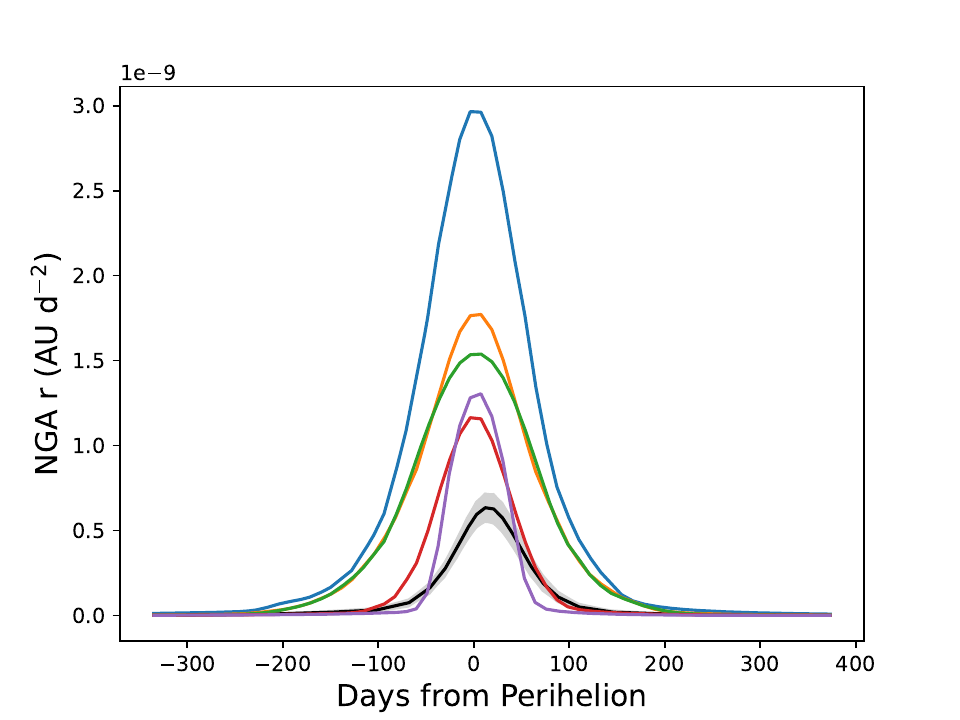}}%
\qquad
\subfloat{\includegraphics[scale=0.5]{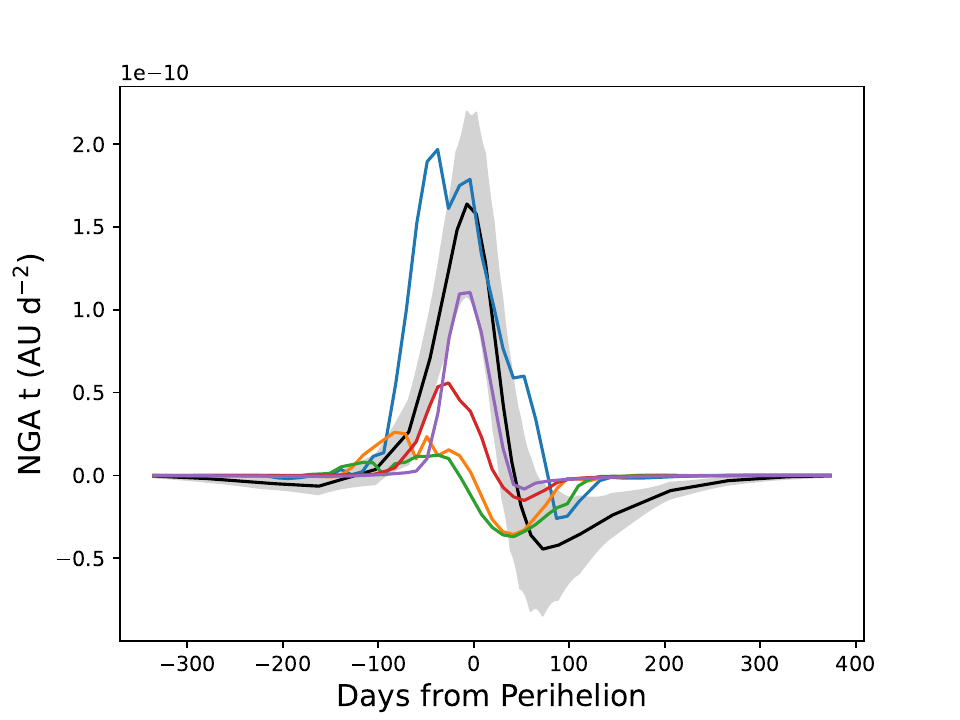}}%
\qquad
\subfloat{\includegraphics[scale=0.5]{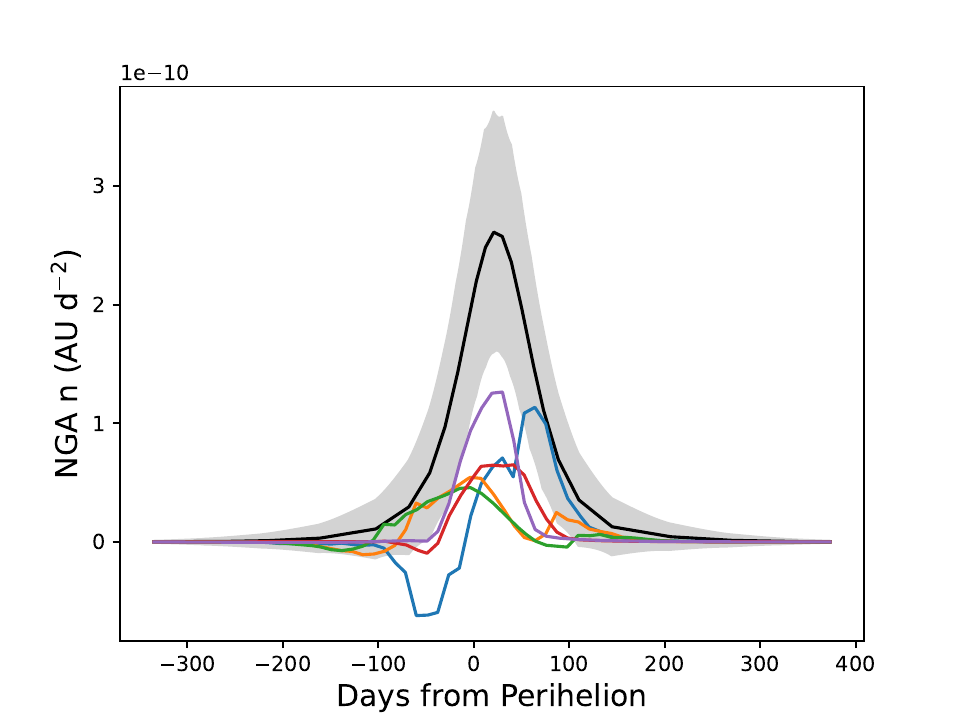}}%
\qquad
\subfloat{\includegraphics[scale=0.5]{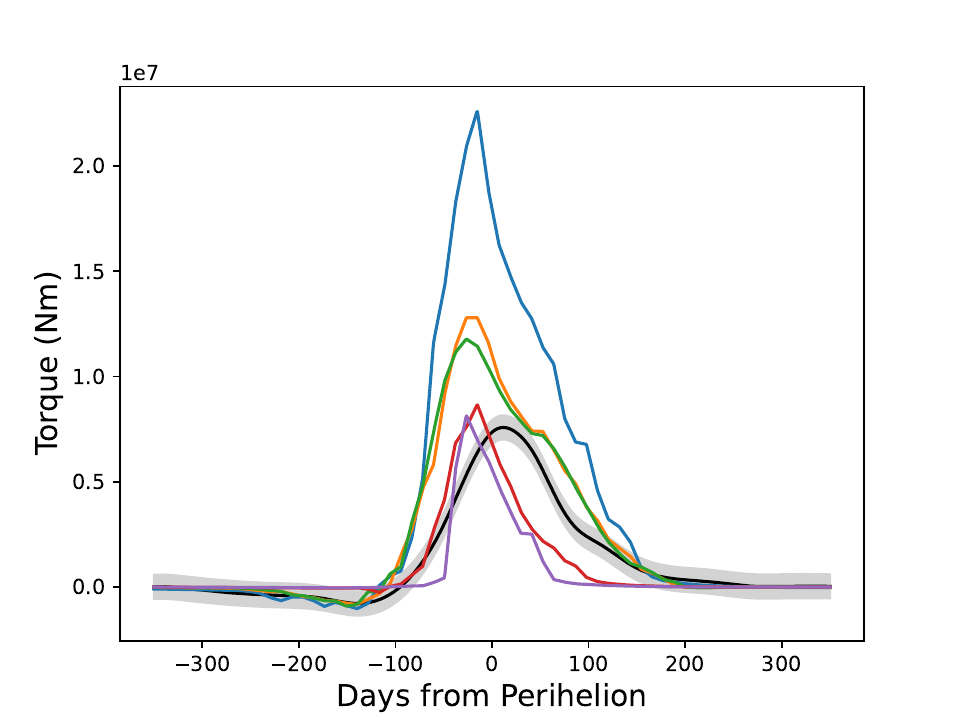}}%
\qquad
\subfloat{\includegraphics[scale=0.5]{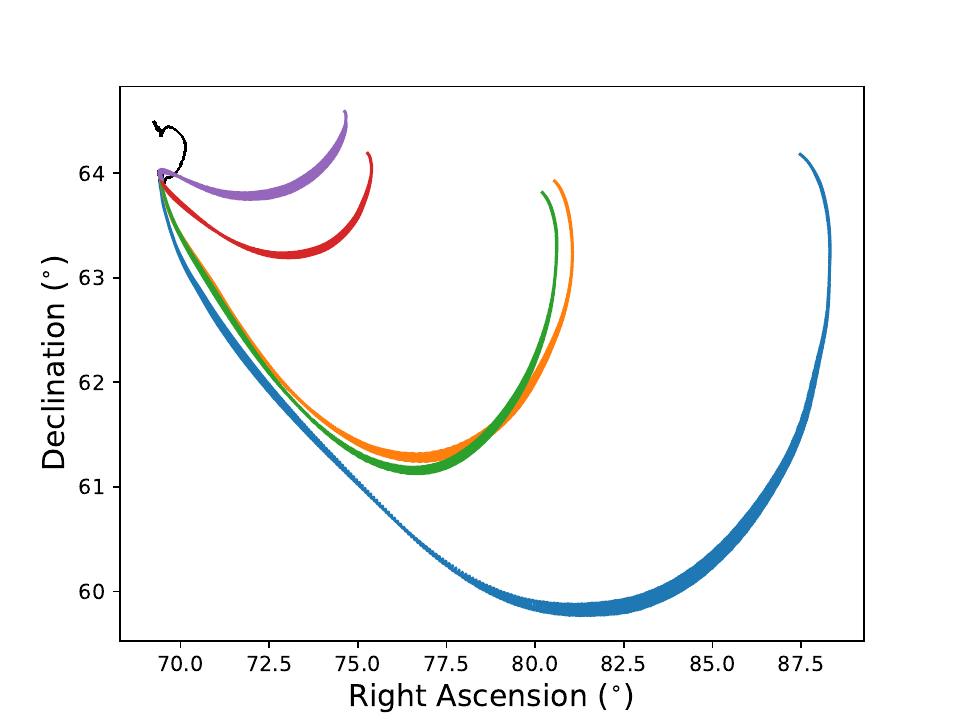}}%
\caption{Results of different model runs with varying parameters compared to the observations (shown by black lines with grey error bounds, and blue points at top left), from top left: total water production rate \citep{Laeuter2020}, radial NGA component, tangential NGA component, normal NGA component \citep{Bologna}, z component of the torque, and pole-orientation (see \citealp{Attree2018}).}
\label{Plot_results}
\end{figure*}

Figure \ref{Plot_results} shows the full range of model outputs compared with Rosetta observations, including: 67P's total water-production rate \citep{Laeuter2020}, its radial, tangential, and normal NGA components \citep{Bologna}, the NGT Z component (change in spin-rate), and its changing pole-orientation (see \citealp{Attree2018} for details). Variations with input parameter are described in the following subsections, but overall we can say that the two-layer model with ejections does a reasonable job in reproducing the general shape of 67P's water production curve. Table \ref{tab:results} gives the overall reduced $\chi^{2}$ values for each model, where the computation and weightings used are the same as in \citet{Attree2024}. Some models produce roughly one order of magnitude too much water, with approximately the right shape, while other more closely match the observed peak water-production at perihelion and match less-well elsewhere. Modelled torque curves are, again, approximately the correct shape, with some significant variation in magnitude, while NGA components and pole-orientations are harder to match. 

These results are generally robust to variations in numerical setup, e.g.~changes in initial conditions such as start-time (aphelion or the beginning of the Rosetta period, as shown here), initial dust-depth ($d$), or the turning on or off of self-heating. We also tested a transition between the surface energy-balance solution (model A) and two-layer solution (model B) at $d>0$, rather than at one aggregate depth, without significant differences in the results (see 'Surface model transition' row in Table \ref{tab:results}). Finally, we also tested variation in the aggregate-packing porosity within the range described above (again, see Table \ref{tab:results}), which shifted up and down the magnitude of the outgassing, without changing the shape of the curve, or those of the other outputs.

\begin{figure*}
\subfloat{\includegraphics[scale=0.5]{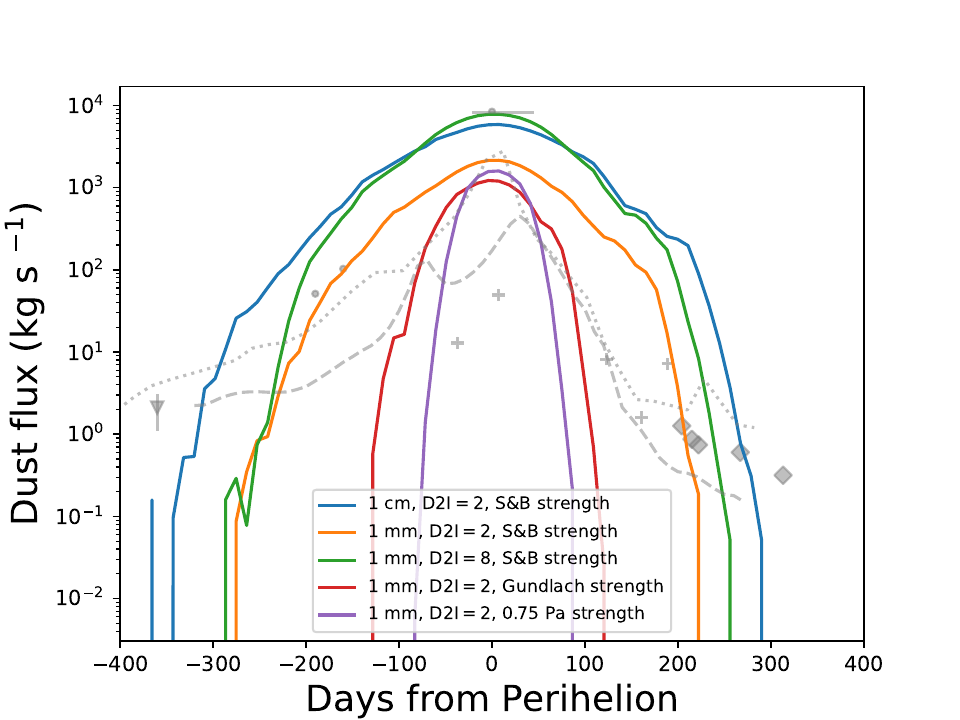}}%
\qquad
\subfloat{\includegraphics[scale=0.5]{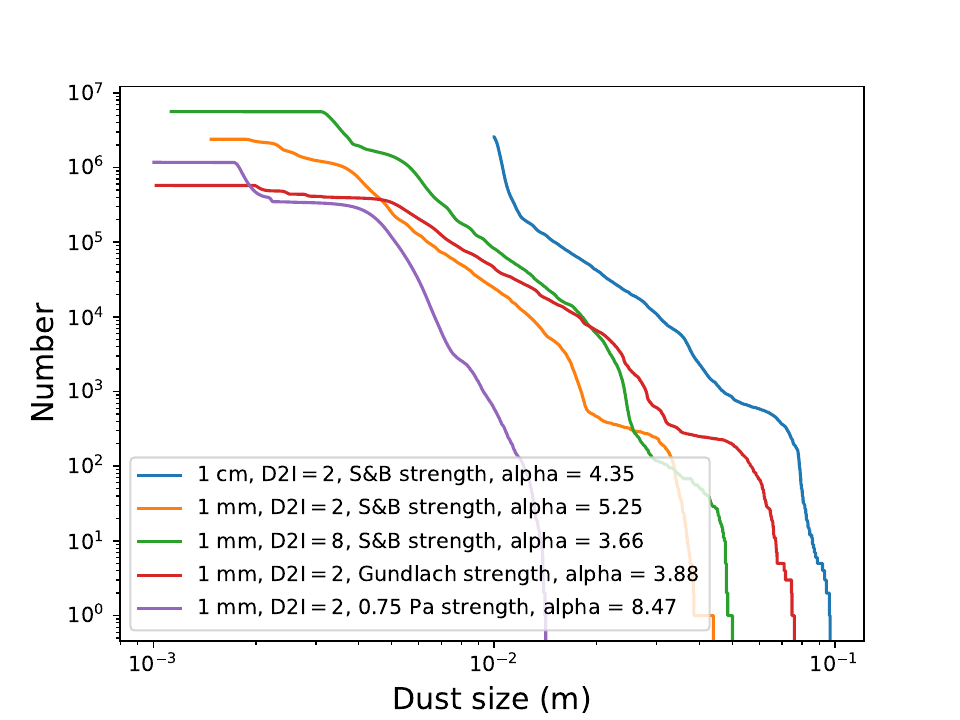}}%
\qquad
\caption{Dust ejection results for different model runs with varying parameters. Left: total launched dust-flux with time, compared to various estimates from Rosetta instruments of launched and ejected dust-fluxes (dashed line: \citealp{Marschall2020} , dotted line: \citealp{Moreno2017}, short solid line: \citealp{Ott2017}, points: \citealp{Fulle2016}, pluses: \citealp{Lemos2024}, diamonds: \citealp{Shi2024}, triangle: \citealp{Rotundi2015}). Right: launched dust size-distribution over the whole period. Fitted power-law exponents are labelled. For comparison, the references generally give exponents between $\alpha\approx3.6-3.8$.}
\label{Plot_results_dust}
\end{figure*}

In addition to the water-production and its resulting NGA and NGT, we can examine the dust that is ejected. Figure \ref{Plot_results_dust} shows the total ejected mass-flux with time, and the overall size-distribution for each of the different model runs. Note that we do not track the dust after ejection, so strictly speaking this is only `launched' dust and we cannot currently say whether it is truly ejected from the comet system, or eventually falls back on the nucleus.

Comparing to a variety of Rosetta and 67P data \citep{Marschall2020, Moreno2017, Ott2017, Lemos2024, Shi2024, Rotundi2015} at different size scales (see \citealp{Attree2025} for further discussion of the observations), and recalling that a significant fraction of launched dust may fall back, we can see that the two-layer model produces roughly the correct order of magnitude of launched dust particles, but struggles to reproduce the shape of the dust emission curves. The model is also limited to ejecting particles between 1 mm and $\sim10$ cm in size, whereas Rosetta detected ejected particles as small as microns or tenths of a micron \citep{Marschall2020}, and as large as tens of centimetres to metres \citep{Shi2024}. The ejections of both small and very large particles are long-standing problems in cometary activity, with additional mechanisms such as internal aggregate pressure build-up \citep{Fulle2019} or separation of grains by sublimation \citep{Schuckart2025}, and CO$_{2}$-pressure \citep{Ciarniello2023, Attree2024b} being invoked.

Another potentially interesting area of difference between the model solutions and observations, is that most of our large particles are actually ejected at the beginning and end of activity, with perihelion ejections dominated by smaller particles. This is in contrast to some observations \citep{Moreno2017}, but is also found in the more complex thermal model of \citet{Attree2025}, where one or more large ejections from a deep sublimation front trigger activity, which then continues with many small, near-surface ejections. Bearing in mind these differences, the agreement of the fitted power-law exponents with the observations ($\approx3.6-3.8$ for the listed references) is reasonable.

We now move on to describe the effects of varying the model input parameters, before returning to comparison with Rosetta observations in Section \ref{sec:discussion}.

\subsection{Aggregate size}
\label{sec:size}

We first vary aggregate size, which directly effects heat- and gas-transport through the dust-crust via the mean-free-path (Eqn.~\ref{eq:MFP}). Strictly speaking, varying aggregate size should additionally change average material strength according to \citet{Skorov.2012}. However, we already take this into account by equating aggregate size with depth in Fig.~\ref{plot_strength}, while we show fixed strengths and different strength-laws separately below in Section \ref{sec:strength} to examine the effects of the different parameters individually.

Large, centimetre-sized aggregates are much more diffusive than millimetre aggregates, allowing higher gas fluxes through the dust-crust and resulting in an overestimated water-production, NGA and NGT as can be seen in the blue versus orange curves in Figure \ref{Plot_results}. Higher rates of emission mean faster dehydration which, combined with the increased radiative thermal conductivity, drives the sublimation front deeper on average than for smaller aggregates. Sublimation fronts for the centimetre aggregates were found to be centimetres to tens of centimetres during the active and inactive phases, respectively, vs.~millimetres to centimetres for the millimetre aggregates. The latter are more compatible with Rosetta observations \citep{Shi2016}.

Deeper sublimation fronts lead to the overall larger particles ejected with the large aggregates, as seen in the Fig.~\ref{Plot_results_dust} size-distribution. Large ejected particles lead to a large dust-mass flux (left-side of Fig.~\ref{Plot_results_dust}), with the weaker material at depth from the depth-dependent strength-law also contributing to more frequent ejections, and an early turning on and long duration of activity (visible in both Figs.~\ref{Plot_results}, and \ref{Plot_results_dust}).

In contrast, smaller aggregates lead to smaller ejected particles, thinner dust-crusts, as well as smaller gas and dust fluxes overall.

\subsection{Dust-to-ice ratio}
\label{sec:D2I}

We next examine the dust-to-ice ratio, $\delta$, which, as described above in Section \ref{sec:method}, only changes the rate of dehydration in the two-layer model. Increasing $\delta$ increases the linear rate of dehydration leading to deeper sublimation fronts on average. This decreases gas fluxes due to damping by the dust-crust, especially at large heliocentric distances where ongoing ejections do not reset the depth. This is visible in Figure \ref{Plot_results} in the green curve compared to orange curve, especially away from perihelion. In tests with very high dust-to-ice ratios ($\delta=50$: \citealp{Ciarniello2023}) this became even more extreme, with very little water production away from perihelion.

Once again, deeper sublimation fronts also lead to slightly larger ejected particles (green versus orange curves on right-side of Fig.~\ref{Plot_results_dust}) as well as a high dust-flux for the large  $\delta$ model. This can be understood by the fast dehydration rate bringing the sublimation front to the weak deep material quickly, resulting in frequent ejection events. This is in contrast to the full numerical models \citep{Attree2025} where thermal inertia in the dust-crust damps some of the energy transport to depth, so that thicker crusts due to faster draining actually reduce activity.

\subsection{Tensile strength}
\label{sec:strength}

Finally, we return to the question of material tensile strength. This parameter is extremely important due to the balance of strength and gas-pressure determining the ejections, which then go-on to control the dust-depth and subsequent outgassing.

Figures \ref{Plot_results} and \ref{Plot_results_dust} red curves show millimetre aggregates with the defects-based depth-dependent strength law used in \citet{Gundlach.2020}, while the purple curves show a fixed value of 0.75 Pa (corresponding to the strength of roughly $d_{agg}\sim2$ mm aggregates, according to \citealp{Skorov.2012}). Both curves show reduced gas- and dust-fluxes compared to the depth-dependence of \citet{Skorov.2012}, due to their overall stronger material acting to damp ejections. Dust ejections are also more limited in time around perihelion, leading to more sharply peaked gas emission curves and a shorter active period. As mentioned above, the first ejection event on a particular facet is often from very deep, meaning that it is the material strength at depth (here, stronger than for the \citealp{Skorov.2012} law) that determines the first ejection and the onset-time of activity.

This complicated relationship between strength, depth, and thermal history is also seen when comparing the ejected size-distributions (right-side of Fig.~\ref{Plot_results_dust}) with the strengths at different depths (Fig.~\ref{plot_strength}). Here, it is illustrative to consider a surface under constant illumination. The sublimation front of such a surface will retreat downwards with a generally decreasing temperature, $T_{ice}$, and corresponding sublimation pressure. The first intersection of this declining pressure with one of the Figure \ref{plot_strength} strength curves will trigger an ejection. If there is never an intersection, because insolation is too low or strength too high, the surface will become inactive and never eject dust. This will often be the case for a fixed, high strength, whereas decreasing strengths with depth give more chance for later, deep ejections. Thus, both depth-dependent strength laws eject more large particles, and more particles overall than the high, fixed strength case. Meanwhile, many ejections occur in the orange curve around sizes of a few millimetres, when compared to the red \citet{Gundlach.2020} curve, whilst there are fewer centimetre-sized ejections. This is because the \citet{Skorov.2012} strength-law is much weaker at depths of a few millimetres, so that our decreasing pressure curve intersects strength here first, rather than deeper. Otherwise-put: weaker near-surface strengths leads to more small ejections, suppressing deeper ejections even with a weak deep material, as we are draining from the surface downwards.

These subtleties demonstrate the importance of tensile strength to the ejection process, especially how it varies with depth, and underlines how our lack of knowledge of this parameter complicates the modelling. We can say, however, that models with \textbf{strongly} decreasing strength with increasing depth are more comparable with the data: they are able to trigger deeper ejections earlier, increasing the active area in space and time. While this would seem to favour the \citet{Skorov.2012} strength-law over that of \citet{Gundlach.2020}, we note that the latter actually shows a better reduced $\chi^{2}$ (Table \ref{tab:results}). This is due to the former's very low strength at shallow depths leading to an overabundance of activity here and very high emission rates. A slightly higher near-surface strength would limit ejections around perihelion, ensuring a smooth ramp-up of activity from the early deep ejections to a slightly damped perihelion activity. This would be in agreement with the findings with the full numerical model \citep{Attree2025}. For these reasons, in the following sections we continue to use the \citet{Skorov.2012} strength-law, alongside a slightly modified version of it with the constant (equal to 1.6 Pa in the nominal case) multiplied by 1.8 (see brown curve in Fig.~\ref{plot_strength}). We note however, that the uncertainties and model-dependencies described here limit our ability to place strong constraints on depth-strength relations.

\subsection{Comparison with previous modelling}
\label{sec:comparison}

\begin{figure*}
\subfloat{\includegraphics[scale=0.5]{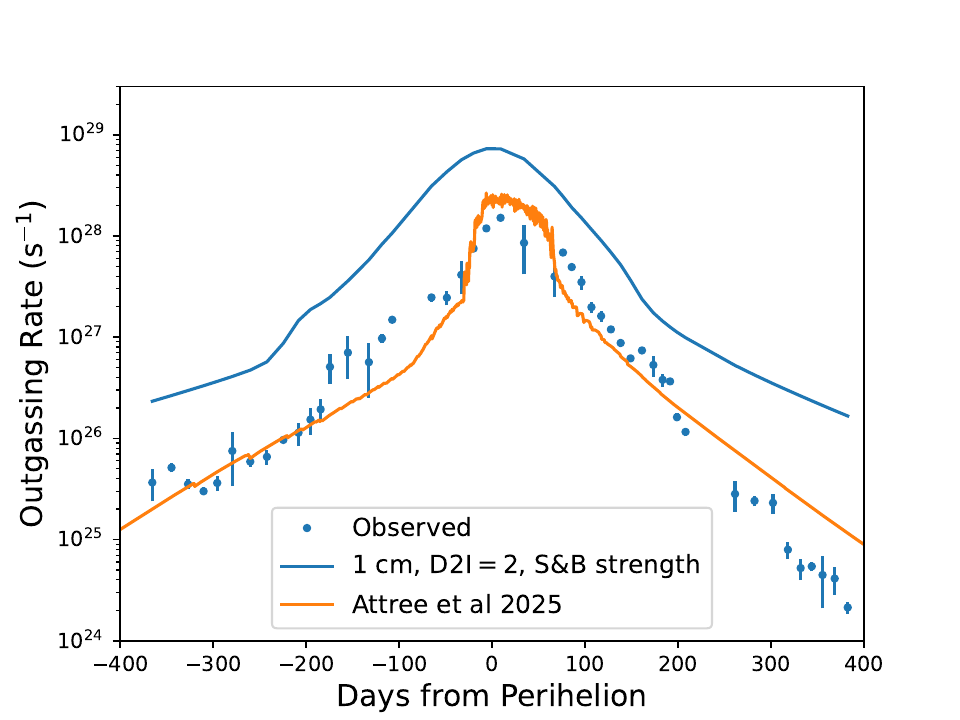}}%
\qquad
\subfloat{\includegraphics[scale=0.5]{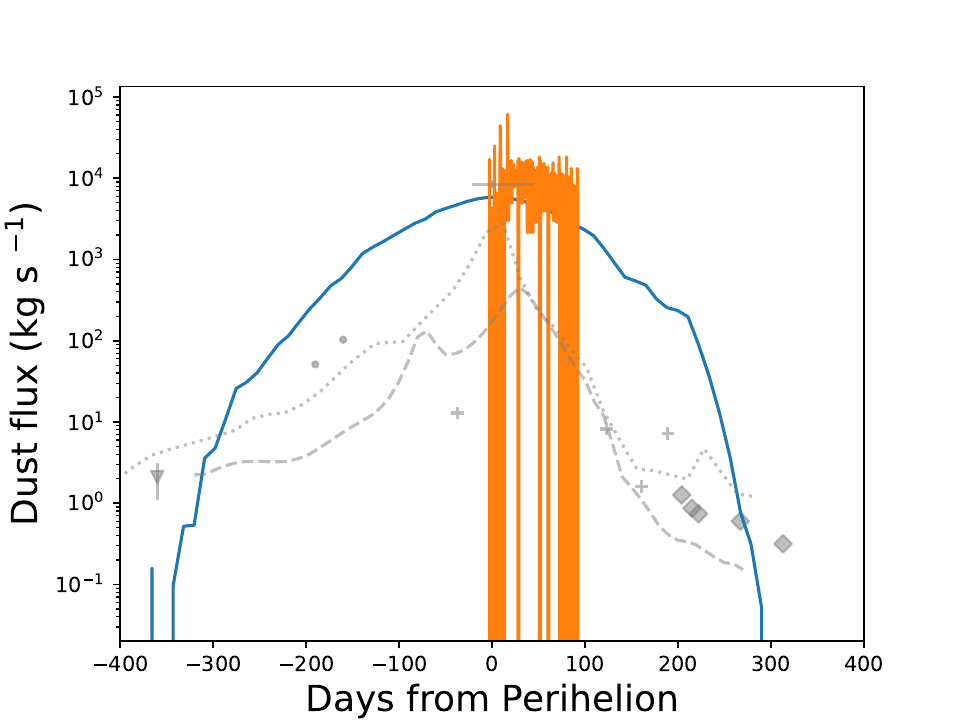}}%
\qquad
\caption{Comparison of this paper's two-layer model (blue curve, labelled input parameters) with the full thermophysical model of \citet{Attree2025} (orange curve, near identical parameters; see text for details). Left: total water production rate with time, Right: total launched dust-flux with time. Discontinuities are numerical noise. Rosetta measurements are also plotted, as described in Figures \ref{Plot_results} and \ref{Plot_results_dust}.}
\label{Plot_comparison}
\end{figure*}

Before continuing, we first pause here to explicitly compare the two-layer model results to our previous, more complex thermophysical model. As detailed in \citet{Attree2025}, this model solves the heat-transfer equation (with an identical treatment of conductivity to here) in a time-dependent manner over many ($\sim500$) layers, whilst approximating the gas flow in a very similar way. In figure \ref{Plot_comparison}, we show the gas production and dust ejections rates of the \citet{Attree2025} model alongside those of the current two-layer model for a similar set of input parameters: $d_{agg}=1$ cm, D2I = 2, $\epsilon=0.4$, $b\approx0.3d_{agg}$. Some small differences remain in the treatment of pressure at the sublimation front between the models, but the primary differences relate to the shape model used (full 67P shape here, spherical nucleus for \citealp{Attree2025}), the spatial resolution (125k facets here, 19 latitude bands for \citealp{Attree2025}), and the presence of a small amount of CO$_{2}$ ice in the complex model ($1\%$ mass relative to water ice shown here); not enough to trigger ejections and hence of limited influence on the behaviour of the water sublimation-front above it.

Aside from the different shape models and CO$_{2}$, we attribute the bulk of the variation between the two curves in Fig.~\ref{Plot_comparison} to the time-resolved nature of the more complex model and the presence of thermal inertia. When compared to the assumed instantaneous equilibration of the two-layer model, the explicit treatment of heat transfer through multiple layers, each with its own internal heat capacity, slows down the propagation of absorbed solar energy into the subsurface and reduces the amount available for sublimation at any given time. This results in lower outgassing rates (left-hand-side) and a reduced dust ejection rate (right-hand-side) as lower sublimation pressures struggle to overcome tensile strengths. A generally lower outgassing was also highlighted in \citet{Attree2024}, when comparing time-resolved results to the instantaneous solution of \citet{Ciarniello2023}. This effect is especially significant at large heliocentric distances, when less absorbed energy goes into sublimation, but here is significant enough to prevent ejections from ever taking place at many latitudes, leaving a much reduced active area in space and time than with the two-layer model. Time-dependent effects also lead to less symmetrical activity curves. Dust thicknesses and ejected particle sizes otherwise remain similar.

\section{Discussion}
\label{sec:discussion}

From the above, we favour small aggregates, with steeply declining tensile strength with depth and a not-too-low water-content, for best agreement between the two-layer model and Rosetta observations. Given all this, however, significant discrepancies still exist between the model curves and the data. To investigate these further, we now qualitatively compare to the empirically-derived water-production rates found for different areas of 67P from combined NGA and NGT modelling \citep{Attree2024} to gain insight into where the more physical two-layer model could be improved.

\begin{figure*}
\subfloat{\includegraphics[scale=0.5]{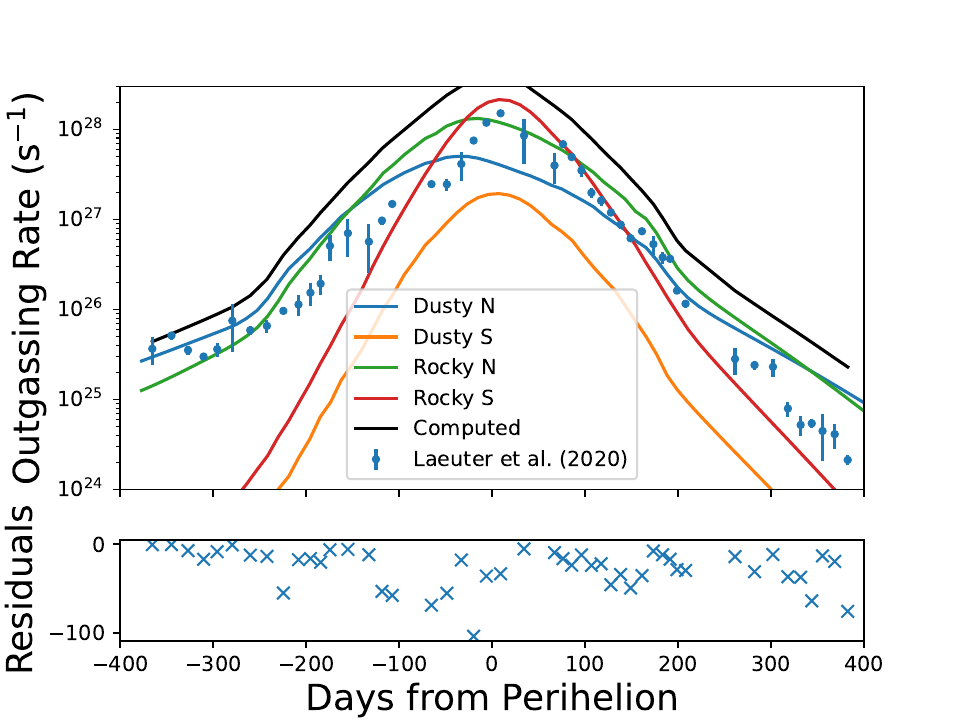}}%
\qquad
\subfloat{\includegraphics[scale=0.5]{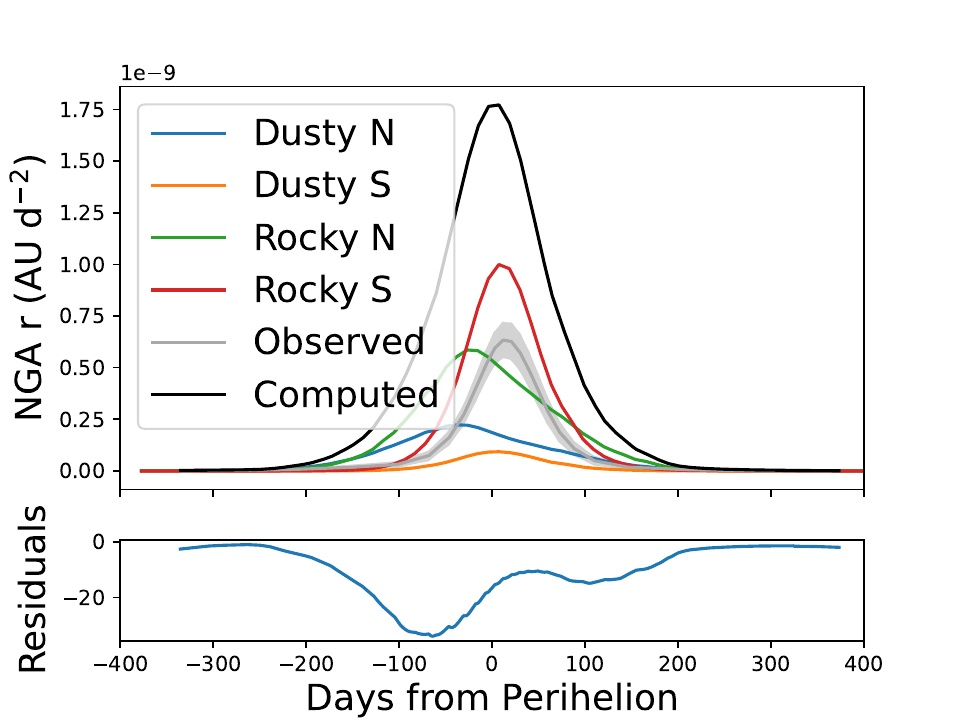}}%
\qquad
\subfloat{\includegraphics[scale=0.5]{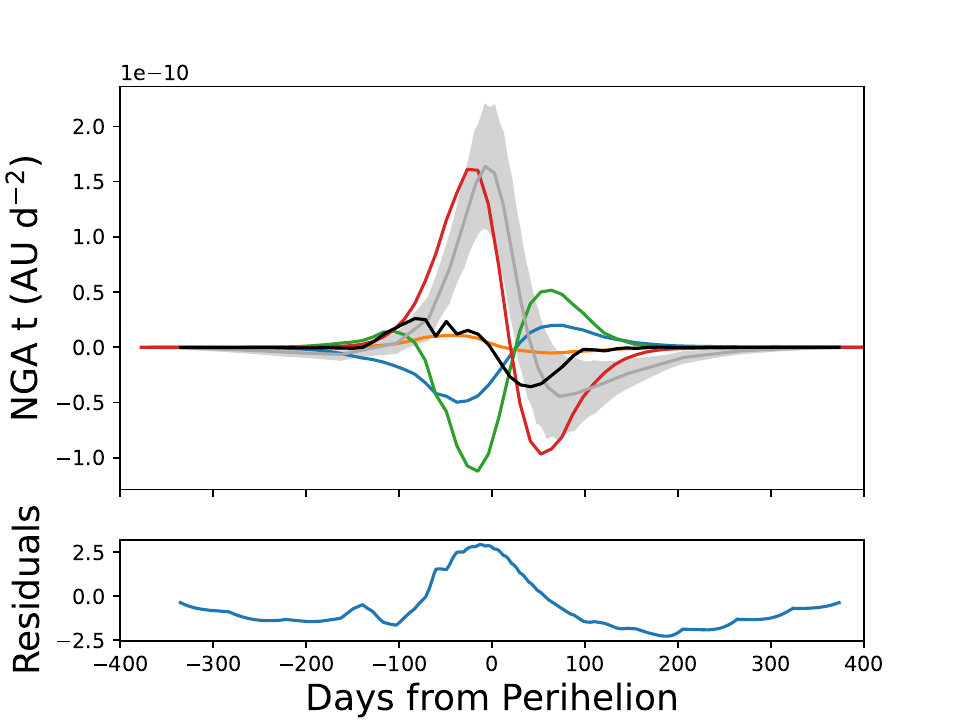}}%
\qquad
\subfloat{\includegraphics[scale=0.5]{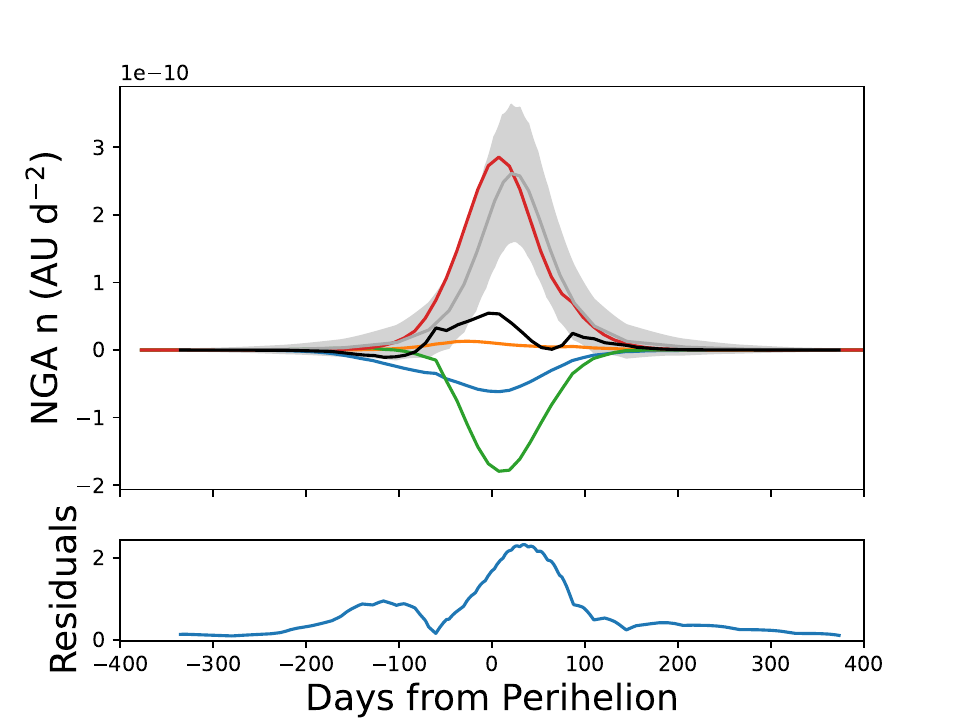}}%
\qquad
\subfloat{\includegraphics[scale=0.5]{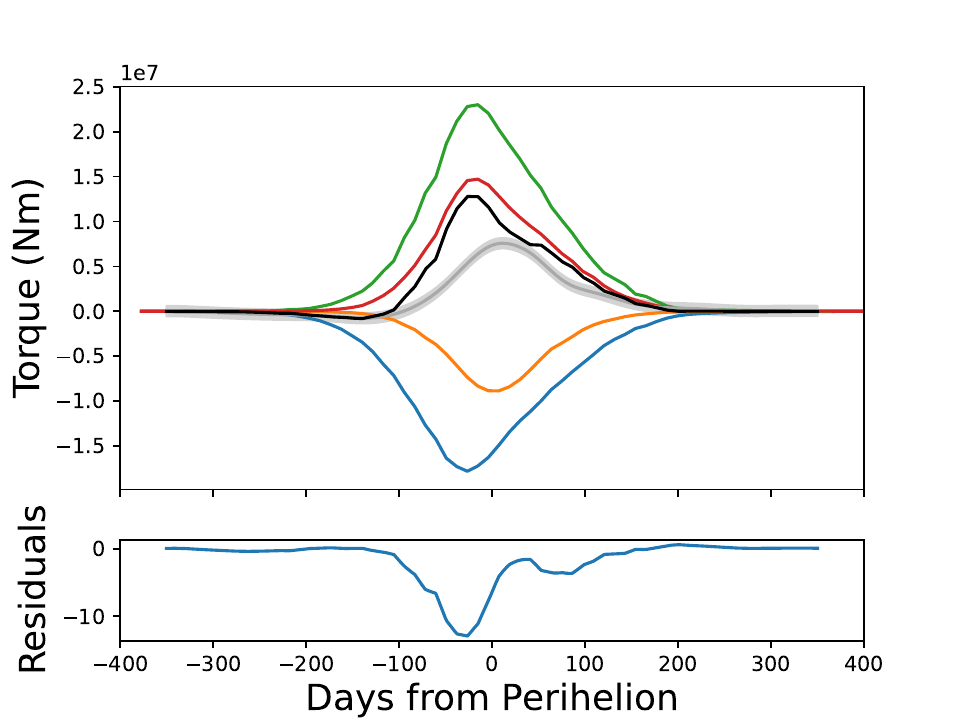}}%
\qquad
\subfloat{\includegraphics[scale=0.5]{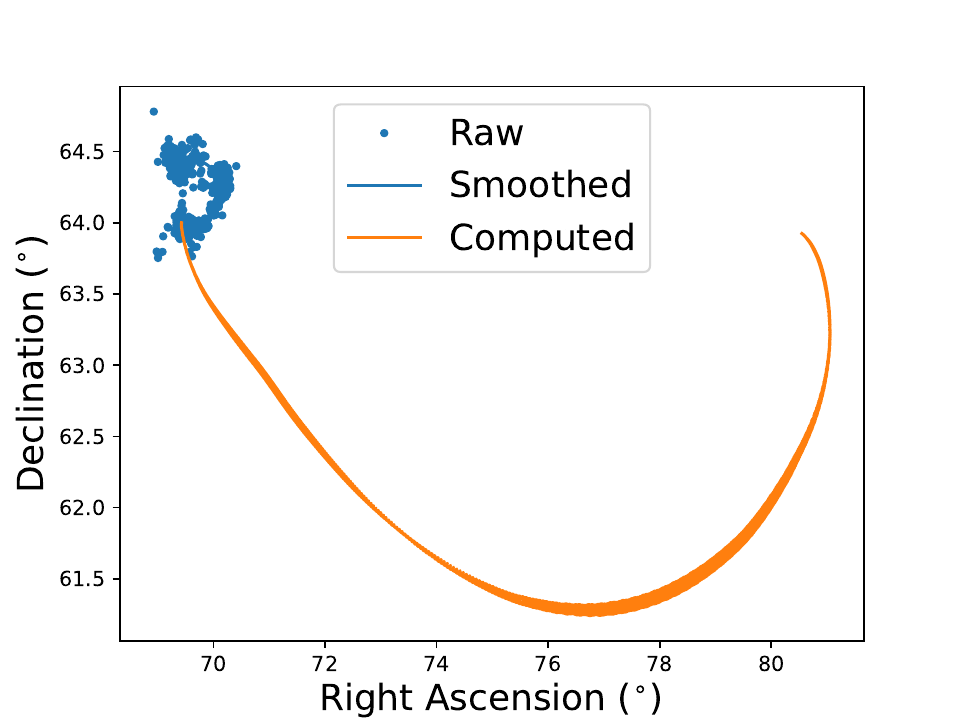}}%
\caption{Same as figure \ref{Plot_results}, but showing a single model run for the case with millimetre pebbles and the \citet{Skorov.2012} strength-law. Individual contributions from the four super-regions of \citealp{Attree2024} are shown alongside the totals (black lines) and the observational data (grey lines and error bounds).}
\label{Plot_results_BestFit}
\end{figure*}

Figure \ref{Plot_results_BestFit} shows the model outputs broken down by region for the case with millimetre aggregates, $\delta=2$, and the \citet{Skorov.2012} strength-law. To recall \citet{Attree2024}, the coloured curves represent the contribution to each total curve (black) of facets grouped by geomorphological unit type \citep{Thomas2018}, and then average orientation, into: rocky north-facing, rocky south-facing, dusty north-facing, and dusty south-facing.

Immediately obvious from this is that the rocky south region (red) contributes to all observations (water-production, all three NGA components, and NGT Z) in the 'right' way, matching the shape of the observed curves well. By contrast, rocky north (green) contributes the `wrong' way to most datasets, especially NGA T and N. For the dusty regions: both south (orange) and north (blue) also mostly contribute the `wrong' way, but the south is limited in its influence on every observable except the Z torque. Here, dusty north can make an important contribution to the initial negative peak seen before $-100$ days to perihelion. Meanwhile, all regions contribute in a highly non-linear manner to the pole-orientation evolution.

Similar patterns are seen when breaking down, by region, the outputs of the runs with different input parameters. Likewise, similar trends for the simplified surface-energy-balance model led the fitting routine in \citet{Attree2024} to try to suppress rocky north by a small effective active fraction (EAF), and to have an EAF that falls-off sharply with time for the dusty north so as to restrict its influence to only the Z torque pre-perihelion. We searched for, but were unable to find, combinations of the variables studied here that would naturally reproduce these trends. Therefore we decided to introduce heterogeneity into the two-layer model by artificially damping ejections in all regions except rocky south for the second orbit, so that their dust depths are controlled solely by dehydration in Eqn.~\ref{eq:d}. For the rocky south, we slightly reduce ejections by using the modified \citet{Skorov.2012} strength-law described above, and shown in Fig.~\ref{plot_strength}. In order to artificially damp ejections in a simple way, without modifying other model parameters, we use an arbitrarily high tensile strength for the other regions. This adjustment could suggest that the rocky south is weaker at the time it is outgassing most around perihelion than other regions on the nucleus are at the times when they would be most active. However, it is important to note that explanations other than increasing the local strength of materials (such as reduced water-content, burying by fallback, or the damping by thermal inertia described in Sect.~\ref{sec:comparison}) could also explain the relative lack of activity in these regions. Those methods are functionally similar to what we propose here, which is merely some form of heterogeneity to damp ejections in all regions outside the rocky south. 

\begin{figure*}
\subfloat{\includegraphics[scale=0.5]{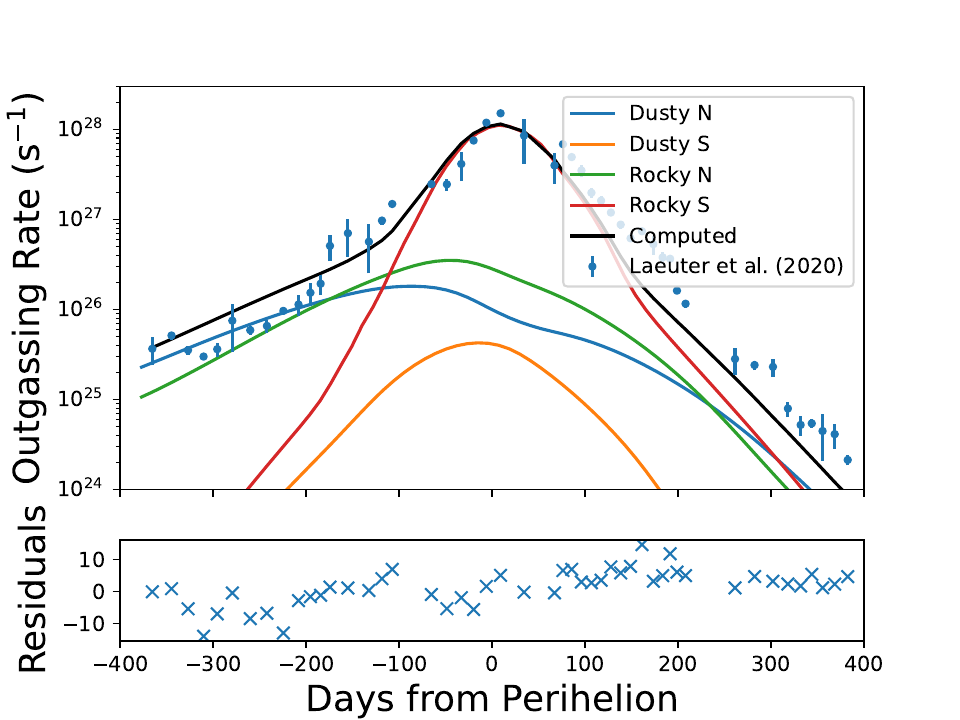}}%
\qquad
\subfloat{\includegraphics[scale=0.5]{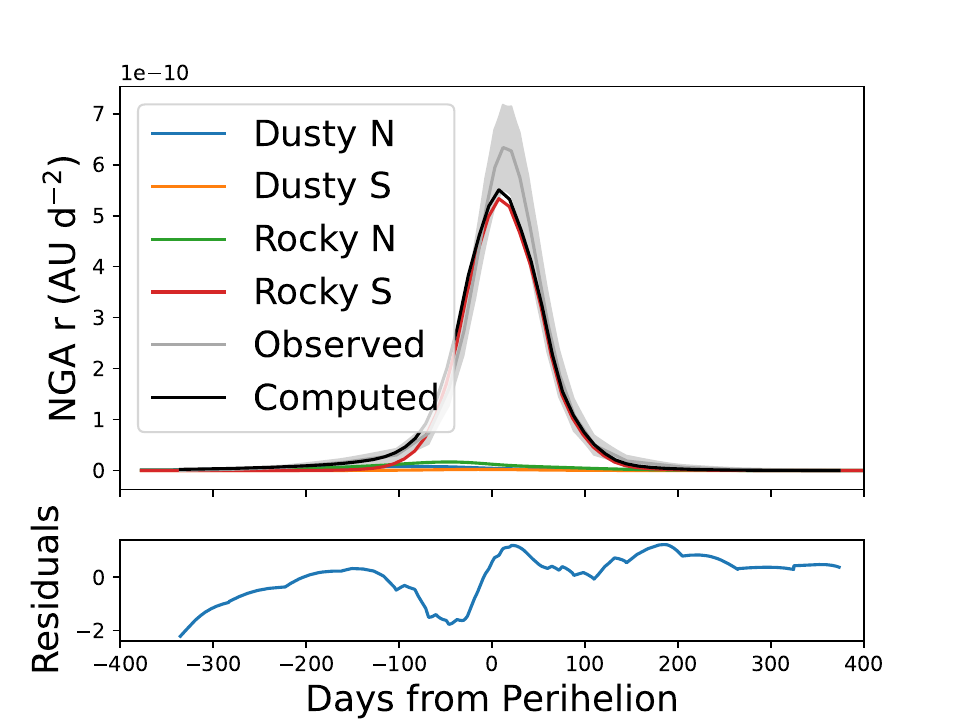}}%
\qquad
\subfloat{\includegraphics[scale=0.5]{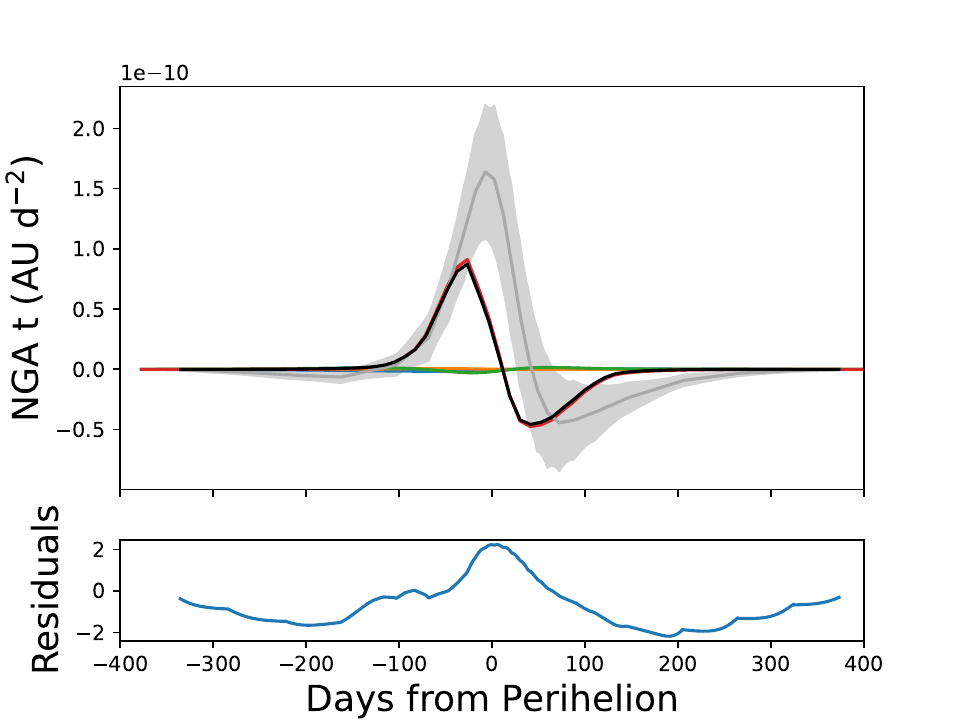}}%
\qquad
\subfloat{\includegraphics[scale=0.5]{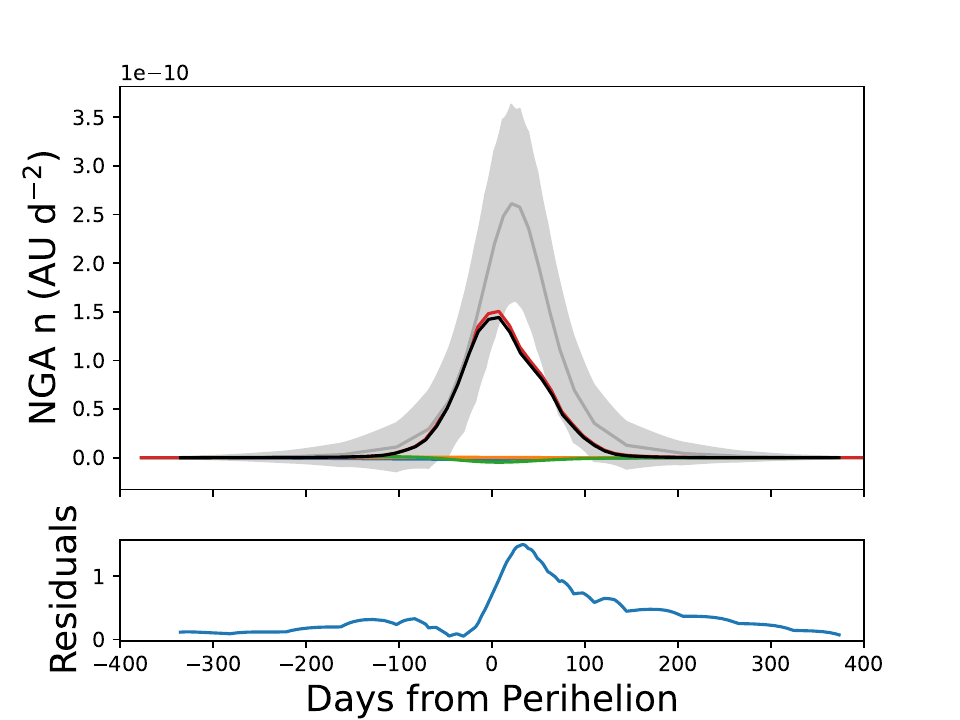}}%
\qquad
\subfloat{\includegraphics[scale=0.5]{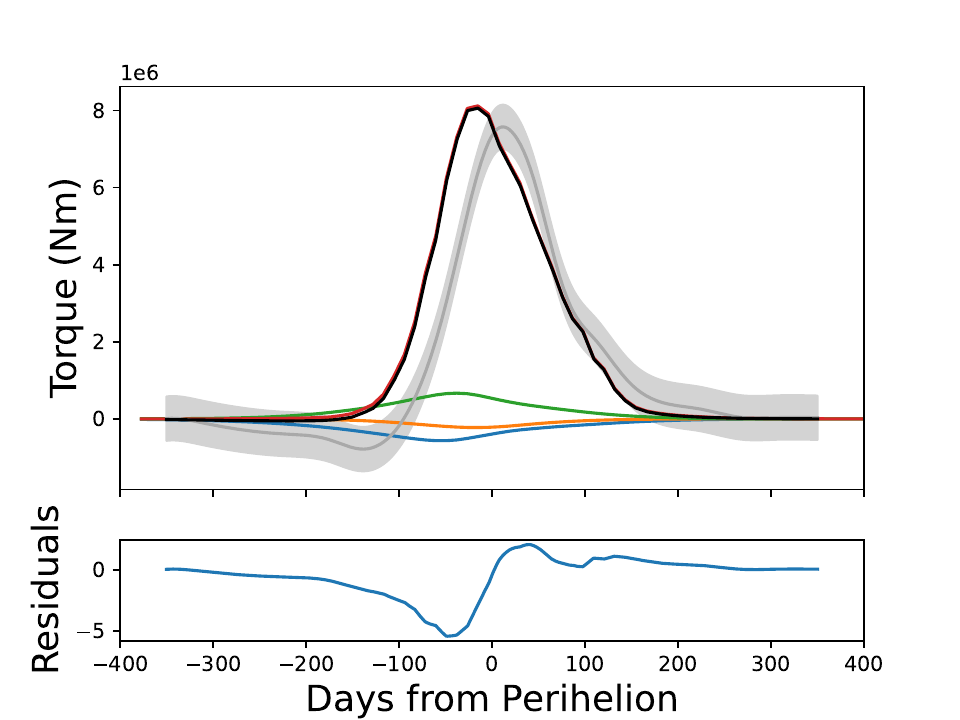}}%
\qquad
\subfloat{\includegraphics[scale=0.5]{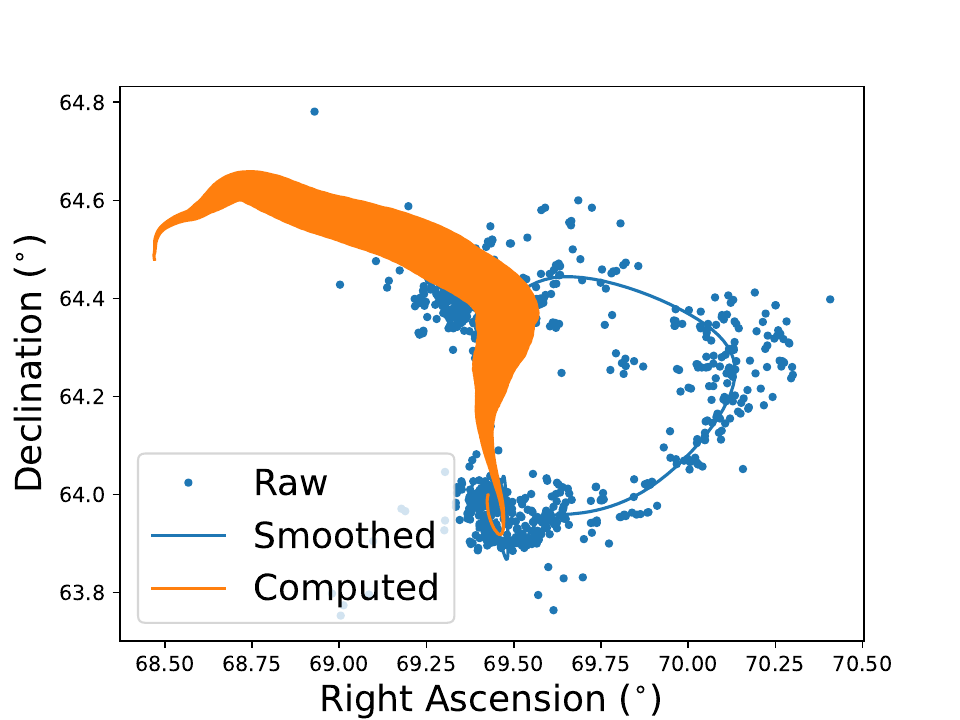}}%
\caption{Same as figure \ref{Plot_results_BestFit}, with a single model run for the case with millimetre pebbles and the \citet{Skorov.2012} strength-law. Individual contributions from the four super-regions of \citealp{Attree2024} are shown alongside the totals (black lines) and the observational data (grey lines and error bounds), where here ejections have been turned off all regions except rocky south (red).}
\label{Plot_results_BetterFit}
\end{figure*}

Figure \ref{Plot_results_BetterFit} shows the outputs broken down by region for this heterogeneous model, with the improvements immediately obvious in the curves as well as the reduced $\chi^{2}$ shown in Table \ref{tab:results}. Water-production, NGA R, T, and N are all now well-matched, while Z torque has the right peak-value and shape, but is still slightly shifted forward by $\sim20$ days relative to the observations. Pole-orientation is also not fully matched by the model; however, declination is reproduced very well, while right ascension is much improved over all the curves shown before (Fig.~\ref{Plot_results}).

The remaining discrepancies in torque could be improved by reducing the early positive torque produced by the rocky south (red) and delaying the overall peak. This could be achieved by delaying the onset of ejections in the south, either by a different strength-depth relation, or the presence of thermal inertia \citep{Attree2025}. Additionally, CO$_{2}$, the outgassing of which becomes significant at and after perihelion but is not simulated here, could help drive more ejections but with a delay caused by its generally deeper sublimation front \citep{Attree2025}. Erosion that happens gradually and sequentially (\citealp{Kossacki_2026} describe a process of removal of grains from the sides of aggregates within the crust, rather than suddenly ejecting the entire crust to the sublimation front, $d$, for example) might also help delay the activity onset in this case, and would help in reproducing comet observations in general \citep{Attree2025}.

Additionally, the early peak in the rocky south can be reduced by compensation from the negative Z torque of the dusty north (blue). Comparing our Fig.~\ref{Plot_results_BetterFit} to Figures 6 and 8 of \citet{Attree2024}, we can see that higher early outgassing in the dusty north and even lower outgassing in the rocky north, helps match the observed NGT, by compensating the early red peak and by also improving the match with right ascension. Large, early outgassing was indeed detected in the Hapi region \citep{Fougere2016, Marschall2019}, which forms part of blue, and may result from water-rich fallback there. From the NGA and NGT, though, this activity must be quenched later and not `reactivated' when the north is illuminated again after the second solstice (see above, \citealp{Attree2024}, and \citealp{Yang2026}), matching observations and implying a refilling of water-rich material at aphelion rather than around perihelion. Indeed, this has been suggested \citep{Attree2024, Jindal2026}, based on ideas of \citet{Davidsson2022}, with material sourced from the nearby Hathor cliff by thermal fracturing of the cold, brittle surface at aphelion (Davidsson et al., 2026, submitted).

Overall then, we see that the regions most differing from the two-layer model predictions are dusty terrains and all those with an average north-facing orientation, which are exactly those most likely to have been modified by fallback \citep{Keller2017} and potential aphelion processes \citep{Davidsson2022}. Conversely, the rocky south, usually considered to be the most primitive terrain due to the strong erosion \citep{Keller2017, Fulle2020, Davidsson2021MNRAS}, has a behaviour which seems well-explained by an alternating dust-crust and ejections model.

\section{Conclusions}
\label{sec:conclusions}

We have implemented a combined two-layer and surface-energy-balance model for cometary outgassing and dust-ejections on the shape-model of comet 67P/Churyumov-Gerasimenko. The transition between the two models is governed by a dynamically varying dust-crust depth which itself is determined by the balance of dehydration and ejections. This is the first step in constructing a coupled activity-fallback model that will include redeposition into the dynamic dust-depth calculation \citep{Jindal2026}. A number of conclusions can already be drawn from these results.

In particular, the two-layer model with only dehydration and ejections can reproduce much of the Rosetta outgassing and dust-ejection data and non-gravitational dynamics. Best agreement is found when the material is made up of small aggregates ($d_{agg}\sim1$ mm) with not-too-high dust-to-ice ratios ($\delta\sim2-8$) and a low strength ($<1$ Pa) that decreases with depth. Under these conditions, dust-particles with sizes of around a millimetre to a decimetre are ejected, mostly from the southern hemisphere and around perihelion. Larger particles may be ejected by CO$_{2}$ \citep{Ciarniello2023, Attree2024b}, while smaller particles and ejections at large heliocentric distances may involve other mechanisms \citep{Fulle2019, Schuckart2025}.

Best agreement between this model and the data is found for dust-crust ejections confined to the southern rocky terrain (as defined in \citealp{Attree2024}). This means that dust ejections elsewhere (as seen by Rosetta, roughly the whole surface of 67P is active: \citealp{Kramer2016, Marschall2020}) may come from the other mechanisms mentioned above, and be rather limited in total mass released when compared to the perihelion activity simulated here. Heterogeneous activity also suggests heterogeneous material properties (tensile strength, water-content, or thermal inertia, for example) but, without yet modelling the fallback, we cannot currently say if this would be an intrinsic property of 67P's nucleus or an evolution due to mass-transport and processing. The difficulty of the model in fitting the behaviour of northern and fallback-dominated areas suggests the latter, but it is precisely this link with the fallback material that will be examined in follow-on work.

Overall, comet 67P's activity remains complex, with multiple ejection mechanisms possibly at play, and cometary outgassing and dust-release determined by a complicated balance between seasonal insolation variations, material properties such as tensile strength, porosity/diffusivity, etc., and material transport and processing.

\section*{Acknowledgements}

N.A. and P.G. acknowledge financial support from project PID2021-126365NB-C21 (MCI/AEI/FEDER, UE) and from the Severo Ochoa grant CEX2021-001131-S funded by MCI/AEI/10.13039/501100011033. Y.S. was funded by the IAA-CSIC Severo Ochoa excellence award CEX2021-001131-S through its incoming visitor programme, and acknowledges the Deutsche Forschungsgemeinschaft (DFG) for support under grant BL 298/32-1. We thank the anonymous reviewer for useful comments and suggestions that significantly improved the text.

\section*{Data Availability}
The data underlying this article will be shared on reasonable request to the corresponding author.




\bibliographystyle{mnras}
\bibliography{paper_references} 


\bsp	
\label{lastpage}
\end{document}